\documentclass[%
 reprint,
 superscriptaddress,
 amsmath,amssymb,
 aps
]{revtex4-2}

\usepackage{amsmath}
\usepackage{printlen}
\usepackage{graphicx}
\usepackage{dcolumn}
\usepackage{bm}
\usepackage{subfigure}
\usepackage{xcolor}
\usepackage{amssymb}
\usepackage{verbatim}
\usepackage{stackengine}
\setstackgap{S}{1pt}
\usepackage{hyperref}
\usepackage{commath}

\renewcommand{\v}{\boldsymbol}
\newcommand{\vv}{\vec}
\renewcommand{\t}{\mathcal}
\renewcommand{\tt}{\underline}

\newcommand{\el}[1]{\textcolor{green}{}}
\newcommand{\eb}[1]{\textcolor{cyan}{}}

\begin{document}

\title{Vibrational lifetimes and viscoelastic properties of ultrastable glasses}

\author{Jan Grießer}
\affiliation{Department of Microsystems Engineering, University of Freiburg, Georges-K\"ohler-Allee 103, 79110 Freiburg, Germany}
\author{Lars Pastewka}
\email[Corresponding author: ]{lars.pastewka@imtek.uni-freiburg.de}
\affiliation{Department of Microsystems Engineering, University of Freiburg, Georges-K\"ohler-Allee 103, 79110 Freiburg, Germany}
\affiliation{Cluster of Excellence livMatS, Freiburg Center for Interactive Materials and Bioinspired Technologies, University of Freiburg, 79110 Freiburg, Germany}

\date{\today}


\begin{abstract}
Amorphous solids are viscoelastic.
They dissipate energy when deformed at finite rate and finite temperature.
We here use analytic theory and molecular simulations to demonstrate that linear viscoelastic dissipation can be directly related to the static and dynamic properties of the fundamental vibrational excitations of an amorphous system.
We study ultrastable glasses that do not age, i.e. that remain in stable minima of the potential energy surface at finite temperature.
Our simulations show four types of vibrational modes, which differ in spatial localization, similarity to plane waves and vibrational lifetimes.
At frequencies below the Boson peak, the viscoelastic response can be split into contributions from plane-wave and quasilocalized modes.
We derive a parameter-free expression for the viscoelastic storage and loss moduli for both of these modes.
Our results show that the dynamics of microscopic dissipation, in particular the lifetimes of the modes, determine the viscoelastic response only at high frequency.
Quasilocalized modes dominate the linear viscoelastic response at intermediate frequencies below the Boson peak.
\end{abstract}

\maketitle

\section{Introduction}

Solid materials deform in response to external load.
In ideal elastic solids, the energy stored in the deformation is fully recovered once the load is removed and they return to their original state.
Real solids always dissipate energy at finite deformation rates.
Such viscoelastic effects are strong for soft polymeric (rubbery) materials, but they are also present in hard materials such as glasses, alloys or even crystals \citep{ferry_viscoelastic_effects_1980, blanter_internal_friction_2007}. 
Viscoelasticity is responsible for limiting the quality factor (line width) of oscillators, and the design of micro- or nanoelectromechanical oscillators seeks to minimize this effect~\cite{lifshitz2000thermoelastic, houston2002thermoelastic, auciello_are_2007, Luo2007-kk, Sumant2014-pg, arash2015review, fan2019intrinsic}.
The resilience of glasses towards impact or shock is determined by its ability to absorb energy, which requires materials with large viscoelasticity~\cite{turneaure2004compressive, khanolkar2016shock, chen2018metallic, renou2017silica}.
Understanding the atomistic origins of viscoelastic dissipation in hard materials is therefore crucial for the design of mechanical devices and resilient materials.

The microscopic mechanism behind viscoelastic dissipation is thermalization.
While the first law of thermodynamics tells us that energy is conserved, ``dissipation'' of energy is the result of processes that evolve an out-of-equilibrium system towards its thermodynamic equilibrium~\citep{kubo_statistical_physics2,zwanzig_nonequilibrium_2001}.
These processes become apparent when we regard only an (open) subsystem of a larger system, e.g. by integrating out part of the microscopic degrees of freedom that do not belong to the subsystem: The separation into a system of interest and a ``heat bath'' (whose detailed state we do not know) leads to the emergence of frictional forces (potentially with memory) in the equations for excitations of the system of interest~\cite{kubo_statistical_physics2,zwanzig_nonequilibrium_2001}.
For solids, such as crystals or glasses, the natural description of excitations is in the form of vibrational normal modes. 
For a three dimensional solid with $N$ atoms and therefore $3N$ vibrational modes, the canonical view is to regard one of these modes as the system of interest and the rest as the heat bath.

In a linear system, each vibrational mode is fully independent and there is no coupling between mode and bath.
Linear systems do not evolve towards thermodynamic equilibrium and there is no ``dissipation''.
Coupling and hence dissipation emerges solely through nonlinearities in the intermode (or interatomic) interactions.
The time required for a vibrational mode to return to the thermodynamic equilibrium is given by its \emph{vibrational lifetime}.
Vibrational lifetimes can either be obtained via simulations or from perturbation theory~\citep{maradudin1962_anharmonic3,fabian1996anharmonic,mcgaughey2014predicting,mizuno_intermittent_rearrangements_2020}. 
The mode-coupling process is particularly interesting for glassy solids that are intrinsically out-of-equilibrium~\citep{debenedetti2001_supercooled_and_the_glass_transition}. 
Besides excitation of vibrational modes, viscoelastic dissipation is triggered by atomic rearrangements, i.e. transitions between inherent structures~\cite{oligschleger1999collective,vollmayr2004single,ciamarra2016particle,leishangthem2017yielding,mizuno_intermittent_rearrangements_2020}. 
We here focus only on the excitation aspect of viscoelastic dissipation in glasses and study ``ultrastable'' glasses, which remain in the same inherent structure even at temperatures comparable to the glass transition temperature~\cite{ninarello2017models}.

In computer glass models, energy dissipation is studied either by measuring the phase shift between the applied strain and the resulting stress in brute-force simulations~\cite{yu_internal_friction_2014,Samwer2016_Correlation,ranganathan_mechanical_damping_2017,adeyemi_measure_viscoelasticity_2022}, or by relating the dynamics of vibrational normal modes to the macroscopic viscoelastic properties~\cite{lemaitre2006sum,damart2017theory,cui_theory_viscoelastic_2017,palyulin_viscoelastic_polymer_2018,kriuchevskyi_scaling_2020,conyuh_viscoelasticity_2021}.
In the brute-force approach, a sinusoidal strain $\gamma(t) = \gamma_0 \sin\left(\Omega t\right)$ with driving frequency $\Omega$ and amplitude $\gamma_0$ is applied to a representative volume element and the resulting stress $\sigma(t)$ is measured.
The temperature is kept constant during deformation by using a thermostat, which typically adds a viscous damping with a characteristic time constant $\tau_\text{BF}$~\citep{berendsen_thermostat_1984,hoover_thermostat_1985,martyna1994_constant_pressure,schneider_Langevin_1978}.
Fitting the stress response with $\sigma(t) = \sigma_0 \sin\left( \Omega t + \delta \right)$ yields stress amplitude $\sigma_0$ and phase shift $\delta$.
If the macroscopic deformation is pure shear with strain $\gamma(t)$, the complex shear modulus $\hat{G}(\Omega) = G^\prime(\Omega) + i G^{\prime\prime}(\Omega)$ is given by~\citep{blanter_internal_friction_2007}
\begin{equation}
    G^\prime 
    = 
    \sigma_0 / \gamma_0 \cos \delta 
    \quad\text{and}\quad
    G^{\prime\prime}
    = 
    \sigma_0 / \gamma_0 \sin \delta,
    \label{eq:dynamic_moduli}
\end{equation}
where $G^\prime$ is the storage modulus (measure of stored elastic energy) and $G^{\prime\prime}$ is the loss modulus (measure of the amount of dissipated energy).

In the microscopic approach, the movement of atoms around their inherent structure is modeled as driven, damped harmonic Langevin oscillators~\citep{lemaitre2006sum,damart2017theory}.
Prior work has used constant microscopic damping with a wavelength-independent relaxation time $\tau$ in such microscopic determination of viscoelastic properties~\citep{lemaitre2006sum,damart2017theory,palyulin_viscoelastic_polymer_2018,kriuchevskyi_scaling_2020}.
The value of $\tau$ was chosen in such a way that the temperature remained constant during deformation or by examining the results for different choices of damping, lacking an intrinsic physical motivation for the specific choice. 
Constant damping leads to dissipation that is independent of wavelength, and this violates momentum conservation.
In other words, fast modes will be generally underdamped and slow modes will be generally overdamped, as discussed for example in Ref.~\citep{Monti2021}.
As we will show below, a good choice for $\tau$ is a value where the whole finite system is underdamped, in which case the time scales of energy flow to the thermostat and of coupling between phononic excitations of the glass are separated.

In the present work, we lift this restriction and use the actual damping constant (or lifetime) of the respective vibrational normal mode. 
Inspired by the seminal works of Ladd, McGaughey and co-workers~\citep{ladd1986lattice,mcgaughey2004quantitative}, we measure the lifetimes from the decay of energy autocorrelation functions. 
This yields a first-principles theory of viscoelastic dissipation in disordered solids.
It allows us to predict the complex shear modulus $\hat{G}$ as a function of driving frequency $\Omega$ and temperature $T$, as well as to make a statement about the accuracy of assuming constant damping. 

\section{Methods}

\subsection{Polydisperse model and preparation protocol}

We use the polydisperse model glass described in Ref.~\citep{lerner2019mechanical} and only summarize the key parameters and the interaction potential here.
The system consists of $N$ particles of equal mass $M$ and varying size $\sigma_i$.
The size $\sigma_i$ is drawn from a distribution $P(\sigma) \sim \sigma^{-3}$ with $\sigma \in \left[\sigma_\text{min} = 1.0, \sigma_\text{max} = 2.22 \right]$, where we use $\sigma_\text{min}$ as our unit of distance. 
The density of the solid is fixed at $\rho=0.58\,\sigma^{-3}$ and periodic boundary conditions are employed in all three spatial directions.
The interaction is governed by a pair-wise additive~\citep{muser2023interatomic} smoothed inverse power law potential of the form 
\begin{equation*}
    U(r_{IJ}) = \left\{
        \begin{array}{ll}
            \epsilon \left[\left(\frac{\sigma_{IJ}}{r_{IJ}} \right)^{\beta} 
            + 
            \sum\limits_{l=0}^q c_{2l} \left(\frac{r_{IJ}}{\sigma_{IJ}} \right)^{2l} \right]
            &, r_{IJ} \leq r_\text{c} \sigma_{IJ} \\
            0 &, r_{IJ} > r_\text{c} \sigma_{IJ} \\
        \end{array}
        \right.
\end{equation*}
where $\beta$ determines the softness, $r_\text{c}$ is the cutoff radius, $\sigma_{IJ}$ is the pairwise interaction length and $\epsilon$ is an energy scale.
The pairwise interaction length is given by the nonadditive mixing rule 
\begin{equation}
    \sigma_{IJ}
    = 
    \frac{1}{2} 
    (\sigma_I + \sigma_J)
    (1-n_a \vert \sigma_I - \sigma_J \vert)
\end{equation}
with $n_a=0.1$. 
The coefficients $c_{2l}$ in the potential are chosen such that the potential is continuous up to the $q$-th derivative at the cutoff~\cite{muser_interatomic_2022}.
Requiring that the potential and its first $q$ derivatives vanish at the cutoff $r_\text{c}$ fixes the coefficients \citep{lerner2019mechanical},
\begin{equation}
    c_{2l} = \frac{(-1)^{l+1}}{(2q -2l)!! 2l!!} \frac{(\beta + 2q)!!}{(\beta -2)!! (\beta + 2l)} r_c^{-(\beta + 2l)}.
\end{equation}
We set $\beta = 10$ and smooth the potential up to the $q=3$ derivative.
Throughout the paper we employ reduced Lennard-Jones units, where energy is measured in units of $\epsilon$, temperature in $\epsilon/k_\text{B}$, time in $\sqrt{M\sigma_\text{min}^2/\epsilon} \equiv t^*$, frequency in $f^* \equiv (t^*)^{-1}$ and stress in $\epsilon/\sigma^3$.

Glassy configurations are obtained by equilibrating the structure at a parent temperature $T_\text{p}$ using the swap Monte-Carlo algorithm \citep{ninarello2017models, lerner2019mechanical}. 
Our ensembles of glasses consist of 100 individual configurations prepared at either $T_\text{p} = 0.3\,\epsilon/k_\text{B}$ or $T_\text{p} = 0.4\,\epsilon/k_\text{B}$.
System size dependence is studied using configurations with $N=8k$, $N=64k$ and $N=256k$ atoms.
The timestep for the molecular dynamics simulations is determined by running equilibrium microcanonical\,(NVE) simulations and studying the change in total energy.
A timestep of $\Delta t = 0.001\,t^*$ for temperatures $T \geq 10^{-2}\,\epsilon/k_\text{B}$ is sufficient to guarantee energy conservation. 
In order to reduce computational cost, a timestep of $\Delta t=0.003\,t^*$ is used at temperatures lower than $T=10^{-2}\,\epsilon/k_\text{B}$. 
Atomic configurations at finite temperature are equilibrated in the canonical\,(NVT) ensemble using a Nos\'e-Hoover chain thermostat with a relaxation time of $1\,t^*$ \citep{shinoda2004,martyna1994_constant_pressure}.

These model glasses interact via purely repulsive forces.
This means that the final configurations are always under hydrostatic stress.
Therefore care needs to be taken when interpreting the elastic constants~\cite{griesser_elastic_constants_2023}.
We come back to this in the discussion below.

\subsection{\label{sec:vibrational_modes}Vibrational modes}

The ultimate goal is to model viscoelastic energy dissipation triggered by the excitation and subsequent decay of vibrational normal modes. 
These vibrational modes are excited because a time-dependent long-wavelength deformation gives rise to non-affine displacements that have a non-zero projection onto the normal modes~\cite{tanguy_continuum_limit_2002,didonna_nonaffine_2005,lemaitre2006sum}.
This mechanism is not limited to disordered systems but also is relevant even for simple crystals~\citep{blanter_internal_friction_2007,ranganathan_mechanical_damping_2017}.

In the following we denote $3$-vectors by arrows (e.g. $\vv{\chi}_i$) and $3N$-vectors using bold symbols (e.g. $\v{e}_m$).
Similarly, we will use an underline for a $3\times 3$ tensor (e.g. $\tt{H}_{IJ}$) and a calligraphic symbol for a $3N\times 3N$ tensor (e.g. $\t{H}$).
Atoms are identified by capital Roman indices, modes (see below) by lowercase Roman indices and Cartesian components by Greek indices.   
We consider a simulation cell whose shape is described by the cell matrix $\mathring{\tt{h}} \equiv \left(\vv{a}_1,\vv{a}_2,\vv{a}_3 \right)$.
The cell is filled with $N$ atoms residing in a local potential energy minimum at positions $\mathring{\vv{r}}_I$.
We refer to this initial state as ``reference'' and mark all quantities which are defined in this reference by a circle.
For amorphous solids, the set of equilibrium positions $\{\mathring{\vv{r}}_I\}$ is typically called the \emph{inherent structure} \citep{ediger1996supercooled, debenedetti2001_supercooled_and_the_glass_transition}.

In the athermal limit, we distinguish between affine and nonaffine mechanical deformation~\cite{maloney2006amorphous}.
The initial simulation cell $\mathring{\tt{h}}$ and the atomic positions $\mathring{\vv{r}}_I$ are remapped by an affine transformation
\begin{equation}
    \vv{r}_I = \tt{h}\cdot\mathring{\tt{h}}^{-1}\cdot\mathring{\vv{r}}_I
\end{equation}
where $\tt{h}$ and $\vv{r}_I$ are now the ``current'' cell and atomic positions.
For amorphous solids, such an affine transformation does not result in an atomic configuration in equilibrium.
Net ``non-affine'' forces $\v{\Xi}_{\alpha\beta} \eta_{\alpha\beta}$ on the atoms remain, where 
\begin{equation}
    \v{\Xi}_{\alpha\beta}
    =
    -\frac{\partial^2 U}{\partial \eta_{\alpha \beta} \partial \v{r}}
\end{equation}
is a force tangent and $\tt{\eta}$ is the Green-Lagrange strain tensor~\cite{lemaitre2006sum,griesser_elastic_constants_2023}.
Allowing the solid to relax to the closest local minimum leads to a configuration with vanishing non-affine forces.
The atomic displacements $\vv{u}_I$ during relaxation are called non-affine displacements.
We can write the current atomic positions as $\vv{r}_I = \tt{h}\cdot\mathring{\tt{h}}^{-1}\cdot\mathring{\vv{r}}_I+\vv{u}_I$.
The non-affine motion is best visualized by pulling the current atomic positions back to the reference and defining pulled-back non-affine displacements as $\vv{\chi}_I \equiv \mathring{\tt{h}}\cdot\tt{h}^{-1}\cdot \vv{u}_I$.

At small finite temperature, atoms vibrate within their inherent structure.
We express these dynamics as a set of coupled, driven Langevin oscillators in the reference configuration,
\begin{equation}
    M \ddot{\v{\chi}}(t)
    +
    \t{F} \cdot \dot{\v{\chi}}(t) 
    +
    \t{H} \cdot \v{\chi}(t)
    =
    \v{\Xi}_{\alpha\beta} \eta_{\alpha \beta}(t)
    +
    \t{S}\cdot
    \v{\xi}(t).
    \label{eq:langevin-atomic-positions}
\end{equation}
The matrix $\t{H}$ is the Hessian with blocks
\begin{equation}
    \tt{H}_{KL} 
    =
    \frac{\partial^2 U}{\partial \vv{r}_K \partial \vv{r}_L},
\end{equation}
$\t{F}$ is a friction matrix and $\t{S}$ is a matrix with noise amplitudes.
The fluctuation-dissipation theorem is~\cite{Grmela1997-sk}
\begin{equation}
    \t{S}\cdot \t{S}^T=2 k_\text{B} T \t{F}
    \label{eq:fluctuation-dissipation}
\end{equation}
and the vector $\v{\xi}(t)$ contains independent white noise variables, i.e. $\langle\xi_{I\alpha}(t)\xi_{J\beta}(t')\rangle=\delta_{IJ}\delta_{\alpha\beta}\delta(t-t')$.
The dynamics of the Langevin oscillators is driven by the non-affine forces that emerge during cell deformation.

The natural frequencies and eigenvectors are obtained from diagonalization of the Hessian, namely by solving
\begin{equation}
    M \omega_{m}^2 \v{e}_m = \t{H} \v{e}_m.
\end{equation}
Since $\t{H}$ is symmetric, the eigenvectors are orthogonal and we also assume that they are normalized such that $\v{e}_m \cdot \v{e}_l = \delta_{ml}$.
Given $\t{Q} =(\v{e}_1, ..., \v{e}_{3N})$, we can write $\t{H} = M\t{Q}\cdot\t{\hat{H}}\cdot\t{Q}^{-1}$ where $\t{\hat{H}}$ is a diagonal matrix of the corresponding eigenfrequencies $\omega_m^2$.
We can use the normal-mode basis $\v{e}_m$ to define modal displacements $\v{X}=M \t{Q}^{-1}\cdot\v{\chi}$.
The dynamical Eq.~\eqref{eq:langevin-atomic-positions} then becomes
\begin{equation}
    \ddot{\v{X}}(t)
    +
    \t{\hat{F}} \cdot \dot{\v{X}}(t) 
    +
    \t{\hat{H}} \cdot \v{X}(t)
    =
    \v{\hat{\Xi}}_{\alpha\beta} \eta_{\alpha \beta}(t)
    +
    \t{\hat{S}}\cdot
    \v{Z}(t)
    \label{eq:langevin-modes}
\end{equation}
with $\t{F} = M \t{Q}\cdot\t{\hat{F}} \cdot\t{Q}^{-1}$, $\t{S} = M \t{Q}\cdot\t{\hat{S}} \cdot\t{Q}^{-1}$ and $\v{\Xi}_{\alpha\beta} = \t{Q}\cdot\v{\hat{\Xi}}_{\alpha\beta}$.
The random vector $\v{Z}(t)=\t{Q}^{-1}\cdot\v{\xi}(t)$ is just a rotation of independent white noise variables, which itself is a vector of independent white noise variables.

The core assumption of the viscoelastic model, which is often not spelled-out explicity, is that $\t{Q}$ not only diagonalizes $\t{H}$, but also the friction matrix $\t{F}$, and therefore by virtue of Eq.~\eqref{eq:fluctuation-dissipation} also $\t{S}$.
This means, the time evolution of each mode $m$ is given by an independent, driven Langevin oscillator
\begin{equation}
    \ddot{X}_m
    + 
    \tau_m^{-1} \dot{X}_m
    +
    \omega_m^2 X_m
    =
    \hat{\Xi}_{m,\alpha\beta} \eta_{\alpha \beta}
    +
    2k_\text{B} T \tau_m^{-1}
    Z_m,
    \label{eq:EOM_vibrational_modes}
\end{equation}
with vibrational lifetime $\tau_m$.
The inverses of the lifetimes are the entries of the diagonal matrix $\t{\hat{F}}$.
The common treatment in the literature assumes $\tau_m=\tau$, independent of the mode $m$~\cite{lemaitre2006sum,damart2017theory,palyulin_viscoelastic_polymer_2018}.
We here relax this specific assumption by using mode-dependent vibrational lifetimes $\tau_m$, which implicitly include full anharmonic scattering~\cite{fabian1996anharmonic,mizuno_anharmonic_properties_2020}.

\subsection{Characterization of modes}

We characterize $\omega_{m}^2$ and $\v{e}_m$ by the vibrational density of states (VDOS) $g(\omega)$, the participation ratio $P_m$ and the phonon order parameter $O_m$~\citep{laird1991_localized_low-frequency_modes, lerner2016_statistics_low-frequency_modes,mazzacurati1996_low_frequency_atomic_motion, shimada2018anomalous, mizuno2017continuum}.
The VDOS is given by
\begin{equation}
    g(\omega) = \frac{1}{3N-3}\sum_{m=4}^{3N} \delta(\omega -\omega_m)
\end{equation}
where $\delta(\omega -\omega_m)$ is the Dirac delta function and we neglect the three translational modes.

A common measure used to characterize the spatial localization of vibrational modes is the participation ratio, which is defined as~\citep{laird1991_localized_low-frequency_modes, lerner2016_statistics_low-frequency_modes,mazzacurati1996_low_frequency_atomic_motion} 
\begin{equation}
	N P_m = \left[ \sum_{I=1}^N (\vv{e}_{mI} \cdot \vv{e}_{mI})^2 \right]^{-1}.
	\label{eq: def_participation_ratio}
\end{equation}
$NP_m$ is the approximate number of atoms participating in the vibration. 
For extended modes, $P_m$ is of the order of unity, independent of $N$, because all atoms participate equally in the vibration. 
For localized modes, it scales with $1/N$ because only a finite number of atoms participate in the vibration. 

In an elastic medium with translational symmetry, we expect a population of plane wave vibrational modes, the phonons.
We evaluate the similarity of the vibrational excitations to plane waves by computing the phonon order parameter defined in Ref.~\citep{mizuno2017continuum,shimada2018anomalous}. 
The idea is to expand the eigenvectors $\v{e}_m$ in the basis of plane waves $\v{w}_{p}(\vv{q})$. 
The basis expansion is written as 
\begin{equation}
    \v{e}_m 
    =
    \sum_{\vv{q},\sigma} A_{m,p}(\vv{q}) \v{w}_p(\vv{q})
\end{equation}
where $A_{m,p}(\vv{q})$ are the expansion coefficients. 
In this equation, $\vv{q}$ is the wave vector and $p$ denotes the polarization, which is either longitudinal or transversal.
The amplitude of a phonon mode with a certain $(\vv{q}, p)$ for an atom $I$ in the amorphous configuration is given by the Fourier basis
\begin{equation}
    \vv{w}_{I,p}(\vv{q}) = \frac{\vv{n}_{p}(\vv{q})}{\sqrt{N}} \exp{\left(i \vv{q}\cdot \mathring{\vv{r}}_I \right)}
\end{equation}
where $\vv{n}_{p}(\vv{q})$ is the polarisation vector.
The longitudinal polarisation vector is defined as $\vv{n}_\text{L}(\vv{q}) = \vv{q}/\vert \vv{q}\vert $ and the two transversal polarisation vectors need to fulfill $\vv{n}_\text{T1}(\vv{q}) \cdot \vv{q} = \vv{n}_\text{T2}(\vv{q}) \cdot \vv{q} = \vv{n}_\text{T1}(\vv{q}) \cdot \vv{n}_\text{T2}(\vv{q}) = 0$.

The phonon-order parameter is defined as the sum of the projection of the eigenvectors $\v{e}_m$ onto the plane-wave basis $\v{w}_{\vv{q}p}$, which are larger than some threshold $\alpha$:
\begin{equation}
    O_m 
    =
    \sum_{\vv{q}=\vv{q}_\text{min}}^{\vv{q}_\text{max}}
    \sum_p
    \left\vert \v{e}_{m} \cdot \v{w}_{p}^*(\vv{q}) \right\vert^2 
    \theta\left( 
    \left\vert 
    \v{e}_{m} \cdot \v{w}_{p}^*(\vv{q}) \right\vert^2  - \alpha
    \right)
\end{equation}
Here $\theta(x)$ is the Heaviside step function and the star indicates the complex conjugate.
The sum runs over all wavevectors from the smallest $\sim 2\pi/L$ (where $L$ is the linear size of the system) up to a largest wavevector, that is given by the minimal distance between two atoms in the configuration.
In our analysis, we use a threshold value of $\alpha=50/(3N-3)$.
We compared different thresholds but found no influence on the result.

We characterize the anharmonicity of the vibrational modes using higher order derivatives of the potential energy as described in Refs.~\citep{gartner2016_nonlinear_modes, gartner2016_nonlinear_modes_2, kapteijns_soft_spots_2020}. 
The atomic positions are displaced from the inherent structure $\mathring{\vv{r}}_I$ in the direction of the eigenvectors $\v{e}_m$ by a distance $s \v{e}_m$, where $s$ is a scale factor of units length. 
We then expand the potential energy $U(\{\vv{r}_I\})$ in a Taylor series, 
\begin{equation}
     U(s \v{e}_m) - U(0)
     =
     \frac{1}{2} \alpha_m s^2 +  \frac{1}{6} \beta_m s^3+ \frac{1}{24} \gamma_m s^4
\end{equation}
with
\begin{align}
    \alpha_m 
    &=
    \sum_{I,J=1}^N
    \frac{\partial^2 U}{\partial \vv{r}_I \partial \vv{r}_J}
    : \left( \vv{e}_{mI} \otimes \vv{e}_{mJ} \right) \\
    \beta_m 
    &= 
    \sum_{I,J,K=1}^N
    \frac{\partial^3 U}{\partial \vv{r}_I \partial \vv{r}_J \partial \vv{r}_K}
    \Shortstack{. . . } \left( \vv{e}_{mI} \otimes \vv{e}_{mJ} \otimes \vv{e}_{mK}\right) \\
    \gamma_m 
    &= 
    \sum_{I,J,K,L=1}^N
    \frac{\partial^4 U}{\partial \vv{r}_I \partial \vv{r}_J \partial \vv{r}_K \partial \vv{r}_K} 
    \Shortstack{. . . . } \left(\vv{e}_{mI} \otimes \vv{e}_{mJ} \otimes \vv{e}_{mK} \otimes \vv{e}_{mL}\right)
    \label{eq: anharmonicity}
\end{align}
where $\otimes$ is the outer product and $:$, $\Shortstack{. . . }$ and $\Shortstack{. . . .}$ are a double, triple and quadruple contraction over all indices of the tensors of second, third and fourth order, respectively, on both sides of the symbol.
Note that $\alpha_m=M \omega_m^2$.

\subsection{Measuring the vibrational lifetimes}

We measure lifetimes of the vibrational modes in microcanonical (NVE) simulations of our equilibrated glasses.
The computation of lifetimes is based on the assumption that the glass remains in its initial inherent structure, $\{\mathring{\vv{r}}_I(t)\} \equiv \{\mathring{\vv{r}}_I(t=0)\}$. 
Therefore it is necessary to discard those simulations, in which the system evolves towards a new potential energy minimum.
A transition towards a new energy minimum is detected from the difference in the inherent structure. 
We define (see also Ref.~\citep{mizuno_intermittent_rearrangements_2020})
\begin{equation}
    \Delta r_\text{IS}(\Delta t) = |\mathring{\v{r}}(t) - \mathring{\v{r}}(t-\Delta t)|
\end{equation}
and test for jumps in $\Delta r_\text{IS}$.
Atomic rearrangements result in values $\Delta r_\text{IS}(\Delta t) \approx 1-10\,\sigma$, independent of temperature. 
Simulations with $\Delta r_\text{IS} > 10^{-2}\,\sigma$ undergo a configurational transition and we discard these simulations.

For those structures where no atomic rearrangement occurred during the simulation, we compute the lifetimes of the vibrational modes from the decay of the autocorrelation of the total energy~\citep{ladd1986lattice,mcgaughey2004quantitative,mishin2016energy}. 
We write the Hamiltonian of the system without external deformation in the harmonic approximation,
$H_\text{harmonic} = \sum_{m} H_m$
with
\begin{equation}
    H_m(X_m(t), \dot{X}_m(t))
    = 
    \frac{1}{2} 
    \left(\dot{X}_m^2(t) + \omega_m^2 X_m^2(t) \right),
    \label{eq:harmonic-approximation}
\end{equation}
the contribution to the Hamiltonian of mode $m$.
The value of $H_m(t)$ for a specific mode $m$ at a specific time $t$ is computed by projecting the atomic configuration $\v{r}(t)$ from the NVE simulation onto the eigenmodes $\v{e}_m$ to obtain $X_m$.
The contribution $H_m(t)$ to the overall energy is then computed from the harmonic approximation, Eq.~\eqref{eq:harmonic-approximation}.

The lifetimes $\tau_m$ are given by the decay of the normalized total energy autocorrelation function $C_m$~\cite{ladd1986lattice,mcgaughey2004quantitative, mcgaughey2014predicting, mishin2016energy},
\begin{equation}
    C_m(t) = \frac{\langle \delta H_m(t) \delta H_m(0) \rangle}{\langle \delta H(0) \delta H(0) \rangle}
\end{equation}
where $\delta H_m(t) = H_m(t) - \langle H_m \rangle$ is the deviation of the current total energy from its expectation value for mode $m$.
The thermodynamic average $\langle\cdot\rangle$ has to be interpreted as being taken over the basin of the potential energy landscape that belongs to the current inherent structure.
We sample it by running dynamic simulations and ensuring that the inherent structure does not change, as discussed in the previous paragraph.

In the high temperature limit, equipartition of energy holds and it follows that $\langle H \rangle = T\,\epsilon$. 
Prior work has shown (numerically and theoretically) that $C_m(t)$ exhibits an exponential decay $C_m \propto \exp\left(-t/\tau_m\right)$ \citep{bickham1998calculation,mcgaughey2004quantitative,mcgaughey2014predicting,mishin2016energy,mizuno_intermittent_rearrangements_2020}.
For fitting $\tau_m$, we compute a first guess by integrating $C_m(t)$ up to a certain maximal threshold,
\begin{equation}
    \tau_m^\prime
    = 
    \int_0^{t_\text{max}} \dif t
    \,
    \frac{\langle \delta H_m(t) \delta H_m(0) \rangle}{\langle \delta H(0) \delta H(0) \rangle},
\end{equation}
where $t_\text{max}$ is taken as the value where the correlation functions drops below $1/e$.
The approximated lifetimes $\tau_m^\prime$ are used as initial guesses for least-squares fits of $\tau_m$ to $C_m(t)$. 
The data is fitted again up to the time where $C_m(t)$ has dropped to $1/e$.

\subsection{\label{sec:viscoelastic_response} Viscoelastic moduli}
In order to derive an expression for the viscoelastic response, we follow the derivation of Lema$\hat{\text{\i}}$tre et al.~\citep{lemaitre2006sum}.
We consider the variation of the stress $\tt{\Pi}$ from time-dependent box deformation, described by a Lagrange strain $\tt{\eta}(t)$.
We write
\begin{equation}
    \Delta \Pi_{\alpha\beta} (t)
    =
    \frac{1}{\mathring{V}}
    \left(
    \left.
    \frac{\partial U}{\partial \eta_{\alpha\beta}}
    \right|_{\tt{\eta}=\tt{\eta}(t)}
    -
    \left.
    \frac{\partial U}{\partial \eta_{\alpha\beta}}
    \right|_{\tt{\eta}=0}
    \right),
    \label{eq:variation_stress}
\end{equation}
where $\tt{\Pi}$ has the properties of a second Piola-Kirchhoff stress.
We consider only small fluctuations and expand Eq.~\eqref{eq:variation_stress} in a Taylor series up to first order,
\begin{equation}
    \Delta\Pi_{\alpha\beta} (t)
    =
    c^\Pi_{\alpha\beta\mu\nu}
    \eta_{\mu\nu}(t)
    -
    \mathring{V}^{-1}
    \v{\Xi}_{\mu\nu}
    \cdot
    \v{\chi}_{\mu\nu}(t)
    \label{eq:taylor_variation_stress}
\end{equation}
where 
\begin{equation}
    c^\Pi_{\alpha\beta\mu\nu}
    =
    \frac{1}{\mathring{V}}
    \left.\frac{\partial^2 U}{\partial \eta_{\alpha\beta}\eta_{\mu\nu}}\right|_{\tt{\eta}=0}
\end{equation}
is the tensor of elastic constants at constant Piola-Kirchoff stress and without non-affine displacements~\cite{muser2023interatomic,griesser_elastic_constants_2023}

Following the derivation of Sec.~\ref{sec:vibrational_modes}, we project onto the vibrational normal modes to obtain
\begin{equation}
    \Delta\Pi_{\alpha\beta}(t)
    =
    c_{\alpha\beta\mu\nu}
    \eta_{\mu\nu}(t)
    -
    \frac{1}{\mathring{V}}
    \sum_m
    \hat{\Xi}_{m,\alpha\beta} X_m(t).
    \label{eq:mode_taylor_variation_stress}
\end{equation}
Considering only oscillatory deformations, we perform a Fourier transformation of Eq.~\eqref{eq:EOM_vibrational_modes} and Eq.~\eqref{eq:mode_taylor_variation_stress}.
This yields
\begin{equation}
    \Delta\tilde{\Pi}_{\alpha\beta}(\Omega)
    =
    D_{\alpha\beta\mu\nu}(\Omega)
    \tilde{\eta}_{\mu \nu}(\Omega)
\end{equation}
with
\begin{equation}
    D^\Pi_{\alpha\beta\mu\nu}
    =
    c^{\Pi}_{\alpha\beta\mu\nu}
    +
    \frac{1}{\mathring{V}}
    \sum_m
    \frac{\hat{\Xi}_{m,\alpha\beta} \hat{\Xi}_{m,\mu\nu}}{\Omega^2 - i\Omega \tau_m^{-1} - \omega_m^2},
    \label{eq:complex_moduli}
\end{equation}
where a tilde indicates a Fourier-transformed quantity.
This result is an extension of the theory derived by Lema{\^\i}tre and Maloney~\citep{lemaitre2006sum} that includes a mode-dependent lifetime $\tau_m$.
We can write this equation in the more common form for the storage $D^\prime_{\alpha\beta\mu\nu}$ and loss modulus $D^{\prime\prime}_{\alpha\beta\mu\nu}$ as
\begin{align}
    D^{\Pi\prime}_{\alpha\beta\mu\nu}
    &=
    c^\Pi_{\alpha\beta\mu\nu}
    +
    \frac{1}{\mathring{V}}
    \sum_{m}
    \frac{\hat{\Xi}_{m,\alpha\beta} \hat{\Xi}_{m,\mu\nu} (\Omega^2-\omega_m^2)}{\left(\Omega^2 - \omega_m^2\right)^2 + \Omega^2 \tau_m^{-2}}
    \label{eq:storage_modulus}
    \\
    D^{\Pi\prime\prime}_{\alpha\beta\mu\nu}
    &=
    \frac{1}{\mathring{V}}
    \sum_{m}
    \frac{\hat{\Xi}_{m,\alpha\beta} \hat{\Xi}_{m,\mu\nu} \tau_m^{-1}\Omega}{\left(\Omega^2 - \omega_m^2\right)^2 + \Omega^2 \tau_m^{-2}}.
    \label{eq:loss_modulus}
\end{align}
Lema$\hat{\text{\i}}$tre et al. showed that the values of $\hat{\Xi}_{m,\alpha\beta}$ can be considered as independent random variables which are self-averaging quantities~\citep{lemaitre2006sum}.
Therefore, we define the non-affine correlator
\begin{equation}
    \Gamma_{\alpha\beta\mu\nu}(\omega) 
    =
    \frac{1}{3N-3}\sum_{m=4}^{3N} \delta(\omega -\omega_m)
    \hat{\Xi}_{m,\alpha\beta} \hat{\Xi}_{m,\mu\nu}.
    \label{eq:correlator}
\end{equation}
Using the non-affine correlator, we can write the storage and loss modulus as
\begin{align}
    D^{\Pi\prime}_{\alpha\beta\mu\nu}
    &=
    c^\Pi_{\alpha\beta\mu\nu}
    +
    \frac{1}{\mathring{V}}
    \int_0^\infty \dif \omega\,
    \frac{g(\omega)\Gamma_{\alpha\beta\mu\nu}(\omega)(\Omega^2 - \omega^2)}{(\Omega^2 - \omega^2)^2 + \Omega^2 \tau^{-2}(\omega)}
    \label{eq:storage-continuum}
    \\
    D^{\Pi\prime\prime}_{\alpha\beta\mu\nu}
    &=
    \frac{1}{\mathring{V}}
    \int_0^\infty \dif \omega\,
    \frac{g(\omega)\Gamma_{\alpha\beta\mu\nu}(\omega) \tau^{-1}(\omega)\Omega}{(\Omega^2 - \omega^2)^2 + \Omega^2 \tau^{-2}(\omega)}
    \label{eq:loss-continuum}
\end{align}
in the thermodynamic limit of large systems.
In the limit $\Omega\to 0$, the loss modulus vanishes and the storage modulus converges to the static elastic constants,
\begin{equation}
    D^{\Pi\prime}_{\alpha\beta\mu\nu}
    \to
    c^\Pi_{\alpha\beta\mu\nu}
    -
    \frac{1}{\mathring{V}}
    \int_0^\infty \dif \omega\,
    \frac{g(\omega)\Gamma_{\alpha\beta\mu\nu}(\omega)}{\omega^2},
\end{equation}
where the last summand is the correction to the elastic constants from non-affine forces~\cite{lemaitre2006sum,griesser_elastic_constants_2023}.
Conversely, in the limit $\Omega\to\infty$ the loss modulus also vanishes but the storage modulus converges to $D^{\Pi\prime}_{\alpha\beta\mu\nu}\to c^\Pi_{\alpha\beta\mu\nu}$, which are the static elastic constants under the assumption of purely affine deformation.

Equation~\eqref{eq:complex_moduli} is valid for arbitrary deformation modes.
Since amorphous solids behave isotropically, the stress is hydrostatic and our interatomic potential is purely repulsive, we consider only the case of simple shear.
Therefore, we define the mean shear modulus,
\begin{equation}
    G^\Pi(\omega)
    =
    \left[D^\Pi_{1212}(\omega) + D^\Pi_{1313}(\omega) + D^\Pi_{2323}(\omega)\right]/3,
\end{equation}
which is the susceptibility that we discuss throughout this paper.
We will also refer to the shear correlator $\Gamma$ below, which is the mean shear component of the correlator, Eq.~\eqref{eq:correlator}.

\subsection{Brute-force calculations}

The previous section describes the linear response theory for viscoelastic dissipation.
We also compute viscoelastic properties from ``brute-force'' calculations of oscillatory cell deformation.
The cell is subjected to time-dependent simple shear with strain $\gamma_{xy}(t)$, which means we evolve the cell matrix according to
\begin{equation}
    \tt{h}(t)
    =
    \begin{pmatrix}
        1 & \gamma_{xy}(t) & 0 \\
        0 & 1 & 0 \\
        0 & 0 & 1
    \end{pmatrix}
    \cdot\tt{\mathring{h}}
\end{equation}
and affinely remap all atomic positions during each strain increment.
We measure the resulting shear stress $\sigma_{xy}$ from the virial theorem~\cite{Admal2010-lr} and fit amplitude $\sigma_0$ and phase lag $\delta$ to compute the complex shear modulus via Eq.~\eqref{eq:dynamic_moduli} (see also Refs.~\cite{yu_internal_friction_2014,ranganathan_mechanical_damping_2017,damart2017theory,palyulin_viscoelastic_polymer_2018,adeyemi_measure_viscoelasticity_2022}).
For each excitation frequency $\Omega$ we simulate 60 periods.
We neglect the first 10 periods during which the system approaches a dynamical steady-state and average the results for each excitation frequency over the remaining 50 periods and ensembles with 50 independent configurations.
The system is strained to $\gamma_0=1.0\%$ and the temperature is kept constant at $T=10^{-2}$~$\epsilon/k_\text{B}$ using a Langevin thermostat with a relaxation time constant of $\tau_{\text{BF}}= 0.1\,t^*$~\citep{schneider_Langevin_1978}.
This maximal strain is well below the yield point.
We tested that using a strain amplitude of $0.5\%$ or $1.5\%$ does not affect the results.

Care needs to be taken when comparing the microscopic theory to brute-force calculations.
In the former, the stress is defined as the derivative of the potential or free-energy with respect to the Green-Lagrange strain tensor $\tt{\eta}$.
In the latter, we compute the instantaneous stress using the virial theorem, which gives the Cauchy stress~\cite{Admal2010-lr}.
Elastic constants derived from either expression differ by a constant that depends explicitly on the stress.
Since our system is (approximately) hydrostatic, the difference between the shear modulus at constant Cauchy stress $G$ and at constant second Piola-Kirchhoff stress $G^\Pi$ is~\cite{wallace_thermoelasticity_stressed_1967,griesser_elastic_constants_2023}
\begin{equation}
    G = G^\Pi - P,
\end{equation}
where $P=-\sigma_{\alpha\alpha}/3$ is the hydrostatic pressure.
In the following we report just $G$ for both microscopic theory and brute-force calculations.

\section{Results}\label{sec: Simulation results}

\subsection{\label{sec:static}Spatial structure of vibrational modes}

In order to classify vibrational modes in amorphous solids, it is useful to define two characteristic frequencies.
The first one is the frequency of the Boson peak, $\omega_{\text{BP}}$, which is determined from the maximum in the reduced VDOS $g(\omega)/ \omega^2$~\cite{phillips_low_temperature_1981,mizuno2017continuum}.
The second one is the frequency of the lowest phonon mode $\omega_{\text{Ph}}$, which can be obtained from~\cite{ashcroft_solid_state_2022}
\begin{equation}
    \omega_\text{Ph} 
    =
    \frac{2\pi}{L} \sqrt{\frac{G^\prime(\Omega\to 0)}{\rho}},
    \label{eq: phonon_frequency}
\end{equation}
where $\rho$ is the density and $L$ is the linear size of the box.
For the ensemble of configurations with $N=8k$ atoms, $\langle \omega_{\text{BP}} \rangle \approx 3\,f^{*}$ and $\langle \omega_{\text{Ph}} \rangle = 1.23\,f^{*}$.
For the large system size with $N=256k$ atoms, the lowest phonon frequency is at $\langle \omega_{\text{Ph}} \rangle = 0.39\,f^{*}$.

\begin{figure}[t!]
    \includegraphics[]{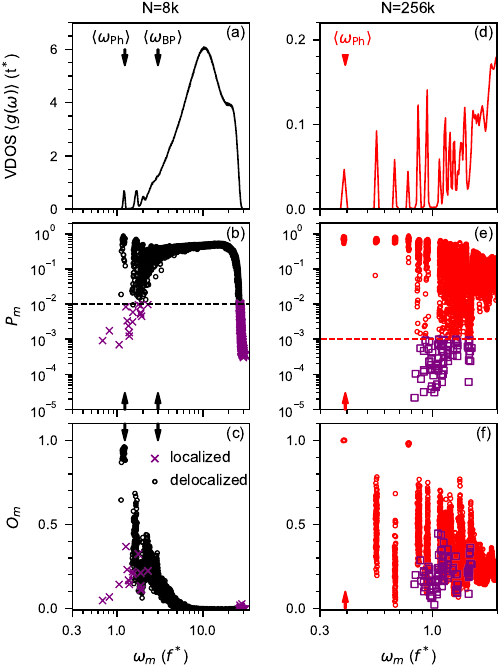}
    \caption{
    \label{fig:static_properties} 
    The ensemble averaged VDOS, the participation ratio $P_m$ and the overlap parameter $O_m$ for a subset of configurations with $N=8k$ atoms (a)-(c) and $N=256k$ atoms (d)-(f).
    Each configuration is prepared at a parent temperature of $T_\text{p}=0.4\,\epsilon/k_\text{B}$.
    The vertical arrows mark the first phonon frequency $\langle \omega_{\text{Ph}} \rangle$ and the location of the Boson peak $\langle \omega_{\text{BP}} \rangle$.
    The horizontal dashed lines mark the thresholds in the participation ratio to distinguish localized and delocalized vibrational modes. 
    Threshold are chosen as $P_m =10^{-2}$ and $P_m =10^{-3}$ for system with $N=8k$ and $N=256k$ atoms. 
    }
\end{figure}

Figure~\ref{fig:static_properties} shows the vibrational density of states (VDOS), the participation ratio $P_m$ and the plane-wave order parameter $O_m$.
At frequencies higher than the Boson peak, we see (Fig.~\ref{fig:static_properties}a) that the VDOS has one dominant local maximum at $\approx 10\,f^{*}$ and one shoulder at $\approx 21\,f^{*}$.
In the mid-frequency range $3\,f^{*} \leq \omega \leq 23\,f^{*}$, we find delocalized vibrational modes (large $P_m$).
In the high-frequency end above $\omega \geq 23\,f^{*}$, we see a sudden decrease of $P_m$, indicative of spatial localization.
The vibrational modes in the mid-frequency as well as in the high-frequency range are poorly described by plane waves, as shown in Fig.~\ref{fig:static_properties}c. 
We call vibrational modes in the mid-frequency range \emph{disordered} and the strongly localized vibrational modes in the end of the frequency spectrum \emph{localized}~\citep{anderson1958absence}.

Vibrational modes with frequencies below the Boson peak show discrete local maxima.
The first peak coincides with the lowest phonon frequency $\omega_{\text{Ph}}$.
Based on the participation ratio and the phonon order parameter in Fig.~\ref{fig:static_properties}b and c, two types of vibrational modes coexist in this limit:
There are clusters of delocalized vibrational modes with similar frequencies $\omega_m$, and large $P_m$ and $O_m$.
The absolute values of $P_m$ and $O_m$ in one of these clusters show large fluctuations.
Based on their similarity to phonons in crystalline materials, we term these vibrational modes \emph{plane-waves}~\cite{mizuno2017continuum,ashcroft_solid_state_2022}.
The second type of vibrational modes observed in the low-frequency range are localized and have a low overlap with plane waves.
In accordance with the literature, we term this type of vibrational mode \emph{quasilocalized}~\citep{laird1991_localized_low-frequency_modes,lerner2016_statistics_low-frequency_modes,mizuno2017continuum}.

Larger systems (Fig.~\ref{fig:static_properties}e-f) show identical behavior.
In the low-frequency limit (Fig.~\ref{fig:static_properties}d), the number of discrete maxima increases over the smaller system as lower vibrational frequencies are resolved.
While the separation between \emph{quasilocalized} and \emph{plane-wave} vibrational modes becomes clearer as a consequence of the increasing system size, the overlap parameter shows larger fluctuations within each cluster of plane-wave modes.
These observations are in agreement with previous works on vibrational modes in disordered solids~\cite{laird1991_localized_low-frequency_modes,lerner2016_statistics_low-frequency_modes,mizuno2017continuum}.

\subsection{\label{sec: lifetimes} Lifetimes of vibrational modes}

Amorphous solids are intrinsically out-of-equilibrium and therefore their atomic structure evolves in time (the structure \emph{ages}) even at temperatures far below the glass transition temperature~\citep{oligschleger1999collective, mizuno_intermittent_rearrangements_2020}. 
In order to suppress aging during measurement of vibrational lifetimes, simulations are performed at a low temperature of $T=10^{-4}\,\epsilon/k_\text{B}$. 
As described in the methods, we detect aging by comparing the atomic positions in the inherent structure and discard simulations when these atomic positions change over time.

\begin{figure*}[ht!]
\includegraphics[]{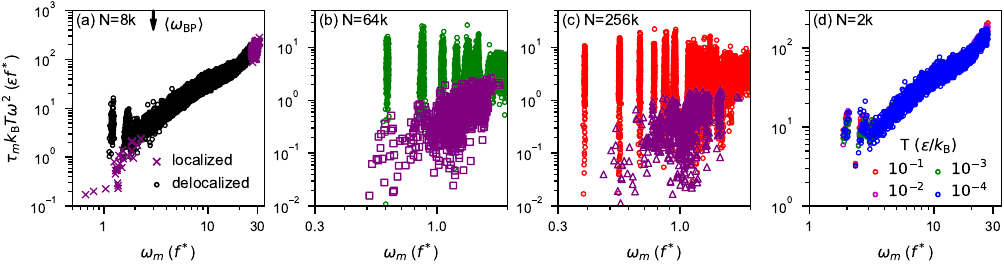}
\caption{\label{fig:vibrational_lifetimes}
         Vibrational lifetimes $\tau_m$ for ten configurations with $N=8k$ atoms (a) and for full ensembles with $N=64k$ atoms (b) and $N=256k$ atoms (c).
         All configurations are prepared at a parent temperature of $k_\text{B}T_\text{p}=0.4\,\epsilon$.
         Lifetimes are computed at a temperature of $T=10^{-4}\,\epsilon/k_\text{B}$. 
         (d) Rescaled vibrational lifetimes for one configuration prepared at $k_\text{B}T_\text{p}=0.3~\epsilon$ with $N=2k$ atoms. The lifetimes are rescaled with $k_\text{B}T\omega^2$ as expected from a third-order perturbation theory. Symbols in each panel are used to distinguish localized and delocalized modes, as determined from their participation ratio.}
\end{figure*}

Figure~\ref{fig:vibrational_lifetimes}a-c show the vibrational lifetimes for configurations with $N=8k$, $N=64k$ and $N=256k$ atoms.
Each ensemble is prepared at a parent temperature of $k_\text{B} T_\text{p}=0.4\,\epsilon$.
For $N=8k$ atoms, we show the full frequency spectrum in Fig.~\ref{fig:vibrational_lifetimes}a.
The high-frequency end, $\omega \geq 23\,f^{*}$, is dominated by localized modes.
Their vibrational lifetimes are frequency-independent.
Decreasing the frequency to $\omega_{\text{BP}}$, increases the vibrational lifetimes.
Below the Boson peak, we identify a splitting of the vibrational lifetimes.
The lifetimes can be separated in vibrational modes with long lifetimes and vibrational modes with short lifetimes.
Comparing this result with Fig.~\ref{fig:static_properties}, we identify the modes with long lifetimes as the plane-wave modes and the latter as the quasilocalized modes.
This splitting of vibrational lifetimes is more clearly visible in Fig.~\ref{fig:vibrational_lifetimes}b, c where we show lifetimes for larger system sizes with $N=64k$ and $N=256k$ atoms.
We can clearly observe that the upper limit of lifetimes of plane-wave modes scales as $\omega^{-2}$ and that their value in a cluster fluctuates over up to two orders of magnitude. 

Aging effects are more pronounced for larger sample sizes and in less stable glasses, which are prepared at higher parent temperature $T_\text{p}$~\cite{mizuno_intermittent_rearrangements_2020}.
To avoid aging, we study the temperature dependence of the vibrational lifetimes for temperatures up to $k_\text{B}T=0.1\,\epsilon$ using amorphous configurations with $N=2k$ atoms prepared at a low parent temperature of $k_\text{B}T_\text{p}=0.3\,\epsilon$.
The highest temperature is in the same order of magnitude as their mode-coupling temperature ($k_\text{B}T_{\text{MCT}} \approx 0.56\,\epsilon$~\cite{lerner2019mechanical,richard_yield_point_2021}).
Third-order perturbation theory predicts the vibrational lifetimes to scale as $\tau_m \propto (k_\text{B}T)^{-1}$~\cite{maradudin1962_anharmonic3,fabian1996anharmonic,mizuno_intermittent_rearrangements_2020}.
Figure~\ref{fig:vibrational_lifetimes}d shows $k_\text{B}T \tau_m$ for one exemplary configuration.
The temperature-scaled lifetimes collapse onto an universal curve.

\subsection{\label{sec:non_affine_correlator} Correlator of non-affine forces}
\begin{figure}[]
    \includegraphics[]{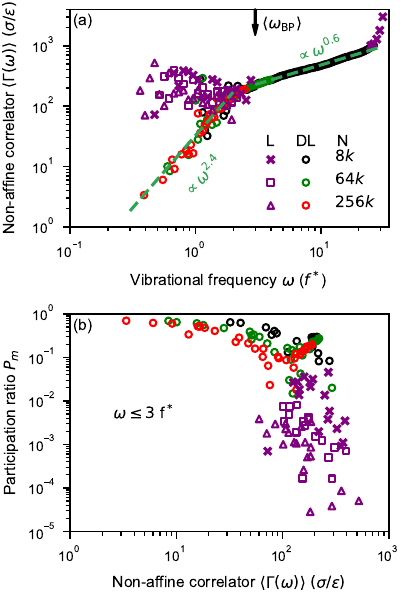}
    \caption{\label{fig:non_affine_correlator}
    (a) Mean shear correlator as a function of vibrational frequency.
    The two green dashed lines are power law fits $\propto c\omega^a$. 
    The power law $\propto \omega^{0.6}$ is a fit to the $N=8k$ data in the frequency range $3 \leq \omega \leq 20$ while the powerlaw fit $\propto \omega^{2.4}$ is fitted to the $N=256k$ data for $\langle \Gamma(\omega)\rangle \leq 10^2$ and $\omega \leq 1.5$.
    (b) Dependence of the shear correlator on the spatial localization of the vibrational modes. Symbols in each panel are used to distinguish localized and delocalized modes, as determined from their participation ratio.
    }
\end{figure}

In Figure~\ref{fig:non_affine_correlator}a we show the ensemble-averaged shear correlator $\langle \Gamma(\omega) \rangle$ defined in Eq.~\eqref{eq:correlator} for configurations with $N=8k$, $N=64k$ and $N=256k$ atoms.
For frequencies above the Boson peak, 
the shear correlator increases with increasing vibrational frequency. 
In the mid-frequency region of the disordered modes, the increase is well represented by a power law $\propto\omega^a$.
By fitting the $N=8k$ data in the frequency range $3\leq \omega \leq 24$ we obtain $a \approx 0.6$.
For vibrational modes with frequencies smaller than the Boson peak frequency, the shear correlator splits in two branches. 
For the first branch we observe a frequency independent shear correlator of $\langle \Gamma(\omega) \rangle \approx 200\,\sigma/\epsilon$.
The shear correlator of the second branch decreases with a power $a \approx 2.4$, computed for the ensemble with $N=256k$ and vibrational frequencies $\omega \leq 1.5\,f^*$.
The behavior in both branches is independent of system size. 
Note that Zaccone and Scossa-Romano~\citep{Zaccone2011_approximate} derived a value of $a = 2$ for an effective medium description of an amorphous solid.

Figure~\ref{fig:non_affine_correlator}b shows the shear correlator as a function of the corresponding participation ratio $\langle P_m \rangle$. 
Spatially delocalized vibrational modes (large $P_m$) have a small shear correlator while localized vibrational modes (small $P_m$) have large values of the shear correlator. 
The shear correlator as a measure of non-affinity is hence also a measure of localization.

\subsection{Frequency-dependent viscoelastic properties}

\begin{figure*}[ht!]
\includegraphics[]{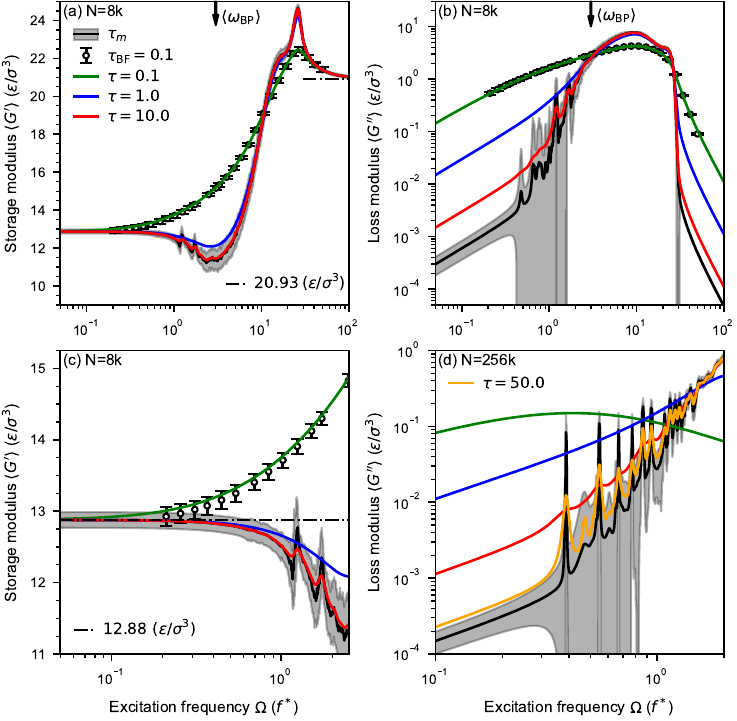}
\caption{
    \label{fig: storage_loss_modulus}
    Ensemble averaged storage $\langle G^\prime \rangle$ (a) and loss modulus $\langle G^{\prime\prime} \rangle$ (b) for the configurations with $N=8k$ atoms in the full frequency range. 
    The vertical arrow marks the location of the Boson peak.
    (c) Detailed view on the storage modulus in the low-frequency limit. 
    The horizontal dashed-dotted line marks the value of $\langle G^\prime \rangle$ in the static limit $\Omega \rightarrow 0$. 
    (d) Loss modulus in the low-frequency limit for the ensemble of configurations with $N=256k$ atoms.
    Results for $\langle G^\prime \rangle$ and $\langle G^{\prime\prime} \rangle$ indicated by lines are computed using Eq.~\ref{eq:complex_moduli}. 
    We employ the actual vibrational lifetimes $\tau_m$ or we assumed a certain constant and mode-independent lifetime $\tau_m = \tau$.
    Results from brute-force simulations are denoted by $\tau_\text{BF}$ and marked by points.
    All lines represent mean values obtained from averaging over multiple configurations while the shaded areas and the error bars mark the standard deviation.
}
\end{figure*}

We compare the frequency-dependent viscoelastic moduli computed by three different approaches.
In the first approach, we compute the shear moduli from Eqs.~\eqref{eq:storage_modulus} and \eqref{eq:loss_modulus} using the mode-dependent vibrational lifetimes $\tau_m$ at a temperature of $T=10^{-2}\,\epsilon/k_\text{B}$.
For the second approach, we assume a constant lifetime i.e. we set $\tau_m = \tau$ and evaluate again Eqs.~\eqref{eq:storage_modulus} and \eqref{eq:loss_modulus}.
In the last approach we perform brute-force (BF) simulations, in which we directly ``wiggle'' the simulation cell and measure the resulting stress (see Methods).
Note that in contrast to the simulations from which we extracted the vibrational lifetimes, we kept atomic trajectories in which atomic rearrangements occurred during mechanical deformation.
The results are summarized in Fig.~\ref{fig: storage_loss_modulus}, which shows storage modulus $G^\prime$ and the loss modulus $G^{\prime \prime}$ for oscillating shear deformation as a function of oscillation frequency $\Omega$.
From Fig.~\ref{fig: storage_loss_modulus}a and b we see that the results for the loss modulus and the storage modulus match perfectly for brute-force calculations and microscopic theory with a constant $\tau_m=\tau_\text{BF}$.
This is an indication that at the temperatures and excitation frequencies studied here atomic rearrangement do not contribute to the viscoelastic modulus.

Having validated the microscopic theory, we now turn to the question of whether an optimal choice of a mode-independent lifetime can recover the correct frequency-dependent behavior.
In contrast to the calculations with $\tau=0.1\,t^*$, the storage modulus has a clear minimum at the Boson peak frequency, $\Omega=\omega_\text{BP}$.
With increasing excitation frequency, the storage modulus increases rapidly and reaches a maximum of $\approx 24.6\,\epsilon/\sigma^3$ at $\Omega \approx 26\,f^{*}$ before dropping to a value of $G^\prime \approx 21\,\epsilon/\sigma^3$ at the high-frequency end.
Using a constant lifetime of $\tau_m = 10\,t^*$ matches the full microscopic theory over the range of excitation frequencies considered here.

Figure~\ref{fig: storage_loss_modulus}c shows the low-frequency limit of the storage modulus.
We present only results for the ensemble of structures with $N=8k$ atoms because the sum in Eq.~\eqref{eq:complex_moduli} requires all eigenmodes up to the Debye frequency of the system for convergence, but this is computationally prohibitive for the larger systems.
The convergence of this sum has been studied in Ref.~\citep{ashwin_cooling_dependence_2013}.
For the small system sizes, the storage modulus in the limit $\Omega \rightarrow 0$ converges independent of the assumed lifetime $\tau$ to the static solution.
The peaks in the storage modulus at low frequencies occur at the characteristic frequencies of the plane-wave vibrational modes.
At the Boson-peak frequency, the storage modulus has a local minimum which can be traced back to a combination of many vibrational modes, long vibrational lifetimes and a high non-affine correlator.

The behavior of the loss modulus Fig.~\ref{fig: storage_loss_modulus}b reflects the shape of the VDOS in Fig.~\ref{fig:static_properties}a, d. 
For increasing excitation frequencies, we observe a dramatic increase of the loss modulus near the Boson peak.
The loss modulus has a broad local maximum at $\Omega \approx 10\,f^{*}$ and a "shoulder" at $\Omega \approx 21\,f^{*}$.
In the low and the high end of the frequency spectrum, the loss modulus decreases rapidly and converges to zero.
Similar to the VDOS, we observe the occurrence of discrete local maxima in the low-frequency end of the spectrum at frequencies of the plane-wave like vibrational modes. 
Below the smallest vibrational frequency in the system, the loss modulus scales linearly with the excitation frequency $\propto \Omega$, see Eq.~\eqref{eq:loss_modulus}.
Examining the results for the loss modulus obtained for constant lifetimes, we see that the mid-frequency and the high-frequency part of the spectrum is in agreement with the correct results as long as $\tau \geq 0.1\,t^*$.
One feature visible in Fig.~\ref{fig: storage_loss_modulus} is that the Boson-peak frequency marks a turning point at which the loss modulus becomes strongly dependent on the choice of lifetime. 
Using fixed lifetimes tend to overestimate the loss modulus below the Boson peak and fails to reproduce the local maxima.
This effect is even more pronounced for larger system sizes shown in Fig.~\ref{fig: storage_loss_modulus}d.

Although we only performed a partial diagonalization, the contribution from vibrational modes with frequencies $\omega_m$ away from the excitation frequency $\Omega$ do not contribute much.
Using mode-dependent lifetimes, we observe that the discrete peaks become narrower with decreasing frequency.
In regions without vibrational modes, the loss modulus is overestimated up to three orders of magnitude while in regions with vibrational modes it tends to be underestimated.

\begin{figure}[ht!]
    \includegraphics[]{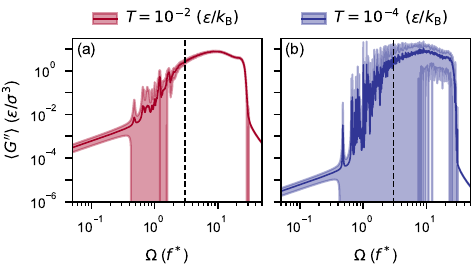}
    \caption{
    \label{fig:loss_modulus_temperature}
    Temperature dependence of the loss modulus $\langle G^{\prime \prime} \rangle$ for the ensemble of configurations with $N=8k$ atoms.
    The vertical dashed line marks the location of the Boson peak.
    In both figures, the straight lines mark the ensemble mean while the shaded areas mark the standard deviation.
    }
\end{figure}

The effect of temperature on the loss modulus is intrinsically included in the vibrational lifetimes.
In Fig.~\ref{fig:loss_modulus_temperature}, we show the loss modulus $G^{\prime\prime}$ for two different temperatures.
The shape and the mean magnitude of the loss modulus above the Boson peak frequency is independent of temperature. 
Below the Boson peak, but in the frequency range where vibrational modes exist, narrower and larger peaks compared to the surrounding are visible for lower temperature. 
It appears, that these discrete peaks have the same amplitude as the mean value at higher temperatures i.e. the peaks are smeared out.
Below the first vibrational mode, the loss modulus decays linearly for both temperatures.
The loss moduli at low frequency differ by a constant factor that depends linearly on temperature.

\section{\label{sec:discussion} Discussion}
The physical interpretation of Eq.~\eqref{eq:complex_moduli} is that storage and loss modulus are determined by the superposition of vibrational modes with eigenfrequencies near the excitation frequency.
Magnitude as well as frequency dependency of the viscoelastic properties are directly related to the properties of the excited vibrational modes.
Based on the excellent agreement in Fig.~\ref{fig: storage_loss_modulus} between brute-force simulations in which atomic rearrangements occur and the theory in Eq.~\eqref{eq:complex_moduli}, excitation of vibrational modes by non-affine displacements is the dominant mechanism of viscoelasticity in the investigated frequency range. 

In the frequency range above the Boson peak, vibrational modes are continuously distributed.
This distribution together with short lifetimes results in a smooth frequency dependency of the viscoelastic moduli due to averaging over many vibrational modes.
The density of vibrational modes is lower towards the high frequency end of the spectrum than near the Boson peak.
These modes have a short lifetime and large non-affine correlator, but the loss modulus at high frequency is low because the low density of states. 

Below the Boson peak, both viscoelastic moduli are dominated by the structure of plane-wave or hybridized modes that occur in discrete frequency bands for the finite-sized systems studied here.
The discrete bands are visible in the loss modulus (Fig.~\ref{fig: storage_loss_modulus}b,d and Fig.~\ref{fig:loss_modulus_temperature}b).
They smooth out at high temperature Fig.~\ref{fig:loss_modulus_temperature}a) as, by virtue of the temperature-dependence of the lifetimes, the line width of each oscillator becomes broader with temperature.
The storage modulus interpolates smoothly between low-frequency and high-frequency limit (see Sec.~\ref{sec:viscoelastic_response}) as each oscillator contributes step-like rather than a Lorentzian function.

Vibrational lifetimes can be directly computed using perturbation theory \citep{maradudin1962_anharmonic3,fabian1996anharmonic,turney2009_pertubation}.
By comparing theoretical and numerical vibrational lifetimes, previous publications showed that cubic anharmonicity is the main source of anharmonic coupling between vibrational excitations in glasses \citep{fabian1996anharmonic,bickham1998calculation,mizuno_intermittent_rearrangements_2020}. 
Furthermore, perturbation theory predicts that vibrational lifetimes scale with temperature $\propto T^{-1}$ which is consistent with our results in Sec.~\ref{sec: lifetimes} \citep{fabian1996anharmonic}.
This scaling is observed to be independent of the type of vibrational mode and even holds for temperatures near the glass transition.

\begin{figure}[ht!]
    \includegraphics[]{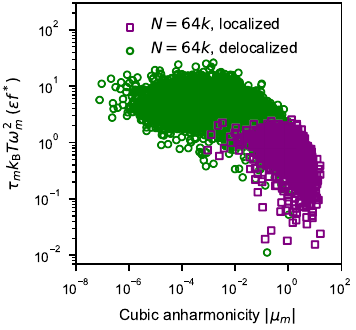}
    \caption{
    \label{fig: anharmonicity}
    Scaled vibrational lifetime as a function of the cubic $\mu_m$ anharmonicity of vibrational modes in the low-frequency limit $\omega \leq \langle \omega_{\text{BP}} \rangle $ for configurations with $N=64k$ atoms.
    }
\end{figure}

To understand whether vibrational modes with shorter lifetimes are related to a larger cubic anharmonicity as suggested by perturbation theory, we evaluate the third order anharmonicity defined in Eq.~\eqref{eq: anharmonicity}. 
In Figure \ref{fig: anharmonicity} we present the anharmonicity of the low-frequency vibrational modes for the ensemble of configurations with $N=8k$ and $N=64k$ atoms prepared at $T_\text{p}=0.4\,\epsilon/k_\text{B}$.
The figure shows that vibrational modes with shorter lifetimes tend to have a larger anharmonicity.
Furthermore, we observe that quasi-localized modes feature larger anharmonicity and therefore shorter lifetimes.

In molecular simulations we can only probe the high-frequency limit of the viscoelastic response, in the regime of THz for real solids.
While the storage modulus appears to converge within the frequency range addressable in the simulation, the asymptotic behavior of the loss modulus remains unclear.
We now attempt to extrapolate the low-frequency behavior of the loss modulus from the general trends of our simulations.
First, we rewrite Eq.~\eqref{eq:loss-continuum} as
\begin{equation}
    G^{\prime\prime}(\Omega)
    =
    \frac{\pi}{\mathring{V}}
    \int_0^\infty \dif \omega\,
    g(\omega)\langle\Gamma(\omega)\rangle
    L(\Omega^2 - \omega^2; \Omega \tau^{-1})
\end{equation}
with normalized Lorentzian $L(x; \epsilon)=\epsilon [x^2 + \epsilon^2]^{-1} /\pi$.
Since the lifetime $\tau$ is roughly constant for quasilocalized modes and scales as $\omega^{-2}$ for plane-waves, i.e. it does not decrease with frequency $\omega$, we can approximate the Lorentzian by a $\delta$-function at small $\Omega$.
This yields
\begin{equation}
    G^{\prime\prime}(\Omega)
    \approx
    \frac{\pi}{2\mathring{V}}
    \frac{g(\Omega) \langle\Gamma(\Omega)\rangle}{\Omega}
\end{equation}
in the low-frequency limit.
This result also means, that using a single relaxation time (as in brute-force calculations) works in the low-frequency response limit because at low enough frequency the dominant modes are always underdamped.
However, studying the high-frequency viscoelastic response near the Boson peak requires either the use of the correct (mode-dependent) relaxation times or of a global relaxation time that is low enough such that the whole system is underdamped (see also Fig.~\ref{fig: storage_loss_modulus}).

Our results and prior work~\cite{biswas1988vibrational,laird1991_localized_low-frequency_modes,oligschleger1993dynamics,lerner2016_statistics_low-frequency_modes, mizuno2017continuum,wang2019low} show that we need to distinguish between delocalized (plane-wave-like) and (quasi-)localized modes in the low-frequency limit.
We therefore split the the loss modulus into these two contributions.
We know that plane-waves have a density of states $g_\text{DL}(\omega)\propto \omega^2$ \citep{debye1912}, while quasilocalized modes contribute with $g_\text{L}(\omega)\propto \omega^4$~\cite{galperin1989localized,buchenau1991anharmonic,baity2015soft,lerner2016_statistics_low-frequency_modes, mizuno2017continuum, wang2019low}
From our data, we find that the correlator $\langle\Gamma(\omega)\rangle$ has power-law behavior $\langle\Gamma_\text{DL}(\omega)\rangle\propto \omega^a$ with $a\approx 2.4$ for plane-waves but is roughly constant $\langle\Gamma_\text{L}\rangle=\text{const.}$ for quasilocalized modes.
We hence find
\begin{equation}
    G^{\prime\prime}_\text{DL}(\Omega) \propto \Omega^{3.4}
    \label{eq:loss_modulus_low_frequency_DL}
\end{equation}
for the delocalized modes and
\begin{equation}
    G^{\prime\prime}_\text{L}(\Omega) \propto \Omega^{3}
    \label{eq:loss_modulus_low_frequency_L}
\end{equation}
for the localized modes.
This result means that the localized modes dominate the low-frequency viscoelastic response.
We note that we cannot rule out from our data that the correlator for the localized modes has a weak power-law dependency $\langle\Gamma_\text{L}\rangle=\Omega^b$.
For $b>a-2\approx 0.4$, the delocalized modes become dominant at low $\Omega$.
Current limitations in accessible frequency are of a numerical nature, since either the diagonalization of the sparse Hessian or the accessible time scales in brute-force calculations become prohibitive for large system sizes.

One question that arises in this context is whether the scaling relations in Eqs.~\eqref{eq:loss_modulus_low_frequency_DL} and  \eqref{eq:loss_modulus_low_frequency_L} are valid for arbitrarily small excitation frequencies in the thermodynamic limit $N\rightarrow \infty$.
This question is directly linked to the existence of localized modes in the thermodynamic limit.
In \citep{bouchbinder_broadening_phonon_bands_2018} it was shown that localized vibrational modes and plane wave-like modes can only be clearly identified and coexist in a frequency range
\begin{equation}
    \omega_g \sim L^{-3/5} \lesssim \omega \lesssim L^{-2/5} \sim \omega_\dagger
\end{equation}
This scaling suggests that in the thermodynamic limit, the two types of mode can no longer be distinguished due to hybridization with each other. 
Another limitation is the existence of two-state processes, which we did not consider in the present work.
As the probability of transition between distinct minima in the potential energy landscape~\citep{oligschleger1999collective,leishangthem2017yielding} of the glass increases with waiting time, these transitions will also become dominant in the low-frequency limit of the loss modulus.
We therefore conclude that our derived scaling represent a distinct intermediate regime, similar to the one observed in wave scattering in glasses \citep{moriel2019wave}.

\section{Summary and conclusion}
In summary, we performed an extensive characterization of vibrational modes in ultrastable glasses.
We observe four different types of vibrational modes which have different static and dynamic properties. 
In the low-frequency limit, where quasilocalized and plane-wave like modes coexist, we observe different scaling relations for vibrational lifetimes of these modes.
While the temperature dependency of both types of vibrational modes agrees with the prediction from perturbation theory, the predicted frequency dependency is observed only for the plane-wave modes.
We used the mode-dependent vibrational lifetimes to extend the theory of Lema$\hat{\i}$tre et al.~\cite{lemaitre2006sum}, yielding a parameter-free theory of linear viscoelasticity. 
The resulting expression indicates, that quasilocalized excitations may become the dominant dissipation channel at intermediate frequency, while the low-frequency viscoelastic response is likely dominated by two-state processes in which the glass jumps between local minima in the potential energy landscape~\citep{oligschleger1999collective,yu_internal_friction_2014,Samwer2016_Correlation,leishangthem2017yielding}.

\section*{Acknowledgements}
We thank Edan Lerner and Eran Bouchbinder for sharing the configurations of ultrastable glasses used in this work as well as for useful comments on the manuscript. 
We thank Wolfram Nöhring and Richard Leute for fruitful discussions.
We used \textsc{matscipy}~\cite{Matscipy} for the calculation of the Hessian, \textsc{Slepc}~\cite{roman2015slepc, hernandez2005slepc} for the solution of large-scale eigenvalue problems and \textsc{LAMMPS}~\cite{THOMPSON2022108171} for the molecular dynamics simulations.
We acknowledge funding by the Deutsche Forschungsgemeinschaft (DFG, grant PA 2023/2).
Calculations were carried on NEMO at the University of Freiburg (DFG grant INST 39/963-1 FUGG) and JUWELS at the Jülich Supercomputing Center (grant hka18).

\providecommand{\noopsort}[1]{}\providecommand{\singleletter}[1]{#1}%

\end{document}